\documentclass[prd, aps, preprint, nofootinbib, 11pt]{revtex4-1}

\usepackage{amsmath}
\usepackage{amssymb}
\usepackage{setspace}
\usepackage{graphicx}
\usepackage[sort&compress]{natbib}
\usepackage{float}
\usepackage[utf8]{inputenc}

\newcommand{\corr}[2]{\ensuremath{\left\langle#1 #2 \right\rangle }}

\begin{document}
 
 %

\begin{center}

\large{\bf A comparison between models of gravity induced decoherence}


\vskip 0.2 in

{\large{\bf Sayantani Bera$^a$\footnote[1]{sayantani.bera@tifr.res.in}, Sandro Donadi$^b$\footnote[2]{sandro.donadi@ts.infn.it}, Kinjalk Lochan$^c$\footnote[3]{kinjalk@iucaa.ernet.in} and Tejinder  P. Singh$^a$}\footnote[4]{tpsingh@tifr.res.in}}

\medskip

{\it $^a$Tata Institute of Fundamental Research,}
{\it Homi Bhabha Road, Mumbai 400005, India}\\
{\it $^b$Department of Physics, University of Trieste, Strada Costiera 11, 34151 Trieste, Italy\\
Istituto Nazionale di Fisica Nucleare, Trieste Section, Via Valerio 2, 34127 Trieste, Italy}\\
{\it $^c$IUCAA, Post Bag 4 , Pune University Campus, Ganeshkhind, Pune 411 007, India.}

\medskip

\end{center}

\centerline{\bf ABSTRACT}
\noindent It has been suggested in the literature that spatial coherence of the wave function can be dynamically suppressed by fluctuations in the spacetime geometry. These fluctuations represent the minimal uncertainty that is present when one probes spacetime geometry with a quantum probe. Two similar models have been proposed, one by Di\'osi [D-model] and one by Karolyhazy and collaborators [K-model], based on apparently unrelated minimal spacetime bounds. The two models arrive at somewhat different expressions for the dependence of the localization coherence length on the mass and size of the quantum object. In the present article we compare and contrast the two models from three aspects: (i) comparison of the spacetime bounds, (ii) method of calculating decoherence time, (iii) comparison of noise correlation. We show that under certain conditions the minimal spacetime bounds in the two models can be derived one from the other. We argue that the methods of calculating the decoherence time are equivalent.  We re-derive the two-point correlation for the fluctuation potential in the K-model, and confirm the earlier result of Di\'osi and Luk\'acs that it is non-white noise, unlike in the D-model, where the corresponding correlation is white noise in time. This seems to be the origin of the different results in the two models.  We  derive the non-Markovian master equation for the K-model. We argue that the minimal spacetime bound cannot predict the noise correlation uniquely, and additional criteria are necessary to accurately determine the effects of gravitationally induced decoherence.

\newpage

\section{Introduction}

\noindent What happens when a quantum system interacts with a classical measuring apparatus?  Why is it that the wave function collapses from being in a superposition of the eigenstates of the measured observable, to being in just one of those states, in violation of the linear superposition principle obeyed by the deterministic Schr\"{o}dinger equation? And what is the origin of the Born probability rule? This set of questions is what is commonly known as the {\it quantum measurement problem}
\cite{Wheeler-Zurek:1983,Bell:87,Albert:92,Leggett:2002,Leggett:2005,Ghirardi:2005,Maudlin:2011}.
 Broadly, there are three classes of explanations which have been investigated in detail.

  The first explanation is to say that collapse never takes place, and to explain the experimental result as a consequence of interaction with the environment, which causes decoherence \cite{Harris:81, Brune:96, Breuer:2000,Joos:03,Schlosshauer:2007, Zeh:70,Caldeira:81,Joos:85,Zurek:03}. This is supplemented with the many-worlds interpretation 
\cite{Everett:57,DeWitt:73,Kent:1990,Deutsch:1998,Vaidman:2002,Wallace:2003,Putnam:05,
Tegmark:2007,Barrett:12,Saunders2010} so that an observer sees only one of the various elements of the diagonalized density matrix.  This explanation requires no change to standard quantum theory, except a reinterpretation (many worlds).
  
The second explanation is Bohmian mechanics \cite{Bohm:52,Bohm2:52,Duerr,Holland,Bohmbook,
Bub:1997,DGZ}  which is a mathematical reformulation of quantum theory, according to which particles move along definite trajectories, but there is a  probability distribution in the initial conditions, which in turn reflects in different outcomes in successive repetitions of a quantum measurement. Bohmian mechanics makes the same experimental predictions as standard quantum theory, as far as both theories are understood.

  The third explanation is that standard quantum theory is an approximation to a stochastic nonlinear quantum theory \cite{Pearle:76,Pearle:79,Pearle:82,Pearle:84,Pearle:89,Pearle:99, 
 Gisin:81,Gisin:84,Gisin:89,Diosi:88a,Diosi:88c,Gisin:95,Weinberg:11,Ghirardi:86}. There is nothing special about quantum measurement; spontaneous collapse of the wave function is an inherent property of the nonlinear theory, but in microscopic systems the collapse occurs so rarely that the theory is effectively indistinguishable from the standard linear Schr\"{o}dinger equation. However, for mesoscopic systems (such as a metal cluster of mass $10^{9}$ amu) and for macroscopic systems (such as the quantum system + measuring apparatus, or  classical objects such as a table or a cat), collapse happens so frequently that any superposition breaks down very rapidly and the system gets well localized in position. The experimental predictions of this nonlinear theory are very close to the standard theory in the micro world, but differ from the standard theory in the meso and the macro world. 
A decisive experiment which chooses between standard quantum theory and stochastic nonlinear theories has not yet been performed, although an extraordinary worldwide effort in this direction is currently in progress \cite{Arndt:2014}. The main reason is that, even if for meso and macro systems the nonlinear effects may be relevant, in this regime the interaction with the environment plays also an important role, which masks the nonlinear effects. 

This is equivalent to saying, and very important to emphasize, that for objects larger than $10^{5}$ amu, quantum theory has not been tested. From $10^{18}$ amu [roughly the scale above which classical mechanics holds], down to $10^{5}$ amu, is an experimentally untested desert which spans some thirteen orders of magnitude!
    
 Thus, the way things stand today, decoherence plus many worlds, Bohmian mechanics, and nonlinear quantum theory, are all valid explanations of observed quantum phenomena and the quantum measurement process. Only future investigations can decide as to which (if any) of these explanations is the correct one; and such investigations are of tremendous importance in helping decide the domain of validity of the standard theory \cite{RMP:2012}.    
 
 One successful formulation\footnote[1]{by successful  we mean the model provides a solution for the quantum measurement problem without violating causality, which is a typical problem of deterministic nonlinear modifications to the Schr\"{o}dinger equation. } of stochastic nonlinear quantum theory, known as Continuous Spontaneous Localization [CSL], was proposed in the eighties \cite{Ghirardi2:90}, and has been studied very extensively since then. Simply put, CSL is a modification of the Schr\"{o}dinger equation, to which a stochastic nonlinear part is added, and two new fundamental constants of nature are introduced, a collapse rate $\lambda$, and a localization length scale $r_c$. CSL explains the collapse of the wave function during a measurement, and it explains the Born probability rule. 

The CSL model is being subjected to stringent experimental tests, and various constraints have also been imposed on its parameters from astrophysical and cosmological observations \cite{RMP:2012}. It is hoped that in the coming decades tests will either verify or rule out this model. Nonetheless, even if CSL were to be experimentally verified, it would still remain a phenomenological model, having been specifically designed to explain collapse of the wave function, and the Born probability rule. Furthermore, the fundamental constants $\lambda$ and $r_c$, as well as their numerical values, have been proposed in an ad hoc manner, so as to be consistent with experimental data. Moreover, the relativistic generalization of CSL is also sought for natural reasons. Significant progress would be made if one could understand why CSL is required in the first place, without taking recourse to the measurement problem, and the Born rule.

Considerable effort has been invested in this direction, and some encouraging results have been obtained. One possibility is to consider CSL as a model which can be derived from a more fundamental underlying theory. A noteworthy effort has been due to Adler and collaborators, where quantum theory, and then CSL model [subject to some specific assumptions] are seen as emerging in the statistical limit from a deterministic theory known as Trace Dynamics \cite{Adler:04}.

Another possibility is to look for some physical mechanism which could be effectively responsible for a significant modification of quantum theory, on macroscopic scales. One mechanism worth considering, and which has been explored in some detail, is the universally present force of gravity. All objects produce a gravitational field, including quantum objects, although today we do not know exactly how the gravitational field of a quantum object is related to its properties such as mass. Since the quantum object does not move on a definite trajectory, the gravitational field produced by it presumably has quantum fluctuations too. The evolution of the object's wave function in such a fluctuating geometry can in principle suffer decoherence and localization, as parts of the wave function that are sufficiently separated in space can lose phase coherence. This principle, or some variation of it, has actually been implemented by a few groups of researchers to show that this effect can cause loss of spatial coherence in macroscopic objects, while having negligible effect on microscopic physics. This phenomenon is commonly referred to as gravity induced decoherence of the wave function.

The first work in this direction was carried out by Karolyhazy and collaborators [we will call this the K-model] \cite{Karolyhazi:66,Karolyhazi:86,Karolyhazy:74, Karolyhazy:90, Karolyhazy:95, Karolyhazy:1982,Frenkel:2002,Frenkel:90,Frenkel:95,Frenkel:97}. Subsequently, gravity induced decoherence models were also developed by Di\'osi [we will call this the D-model] \cite{Diosi:87a,Diosi:87,Diosi:89}.
While we do not discuss a third development here, mention must be made of the important work of Penrose \cite{Penrose:96} on the effect of self-gravity on quantum evolution, and the subsequent investigations by various researchers on the Schr\"{o}dinger-Newton equation, reviewed for instance in \cite{RMP:2012, Bahrami2014a,Singh:2014}.

It is important to stress that in the K-model as well as in the first version of the D-model \cite{Diosi:87a,Diosi:87} the evolution of the state vector is given by a random unitary Schr\"{o}dinger equation i.e. a Schr\"{o}dinger equation where a stochastic potential describes the fluctuations in the geometry of the spacetime. Therefore, in contrast  to the CSL model, these two models do not describe any real collapse of the wave function: they only explain an appearance of decoherence effects due to the presence of the stochastic potential. However, there is an important connection between random unitary Schr\"{o}dinger equations and nonlinear stochastic equations like the one of collapse models. As shown in \cite{Adler:2007}, given a random unitary Schr\"{o}dinger equation, there is always a corresponding nonlinear equation (which has exactly the form as in the collapse models equation), which leads to the same master equation. Therefore, even though the dynamics for the state vectors in these models are very different, as far as we are concerned with averaged quantities, (derivable from the master equation), the two dynamics lead to exactly same predictions. As an example, in the case of the D-model, the corresponding collapse equation was proposed in \cite{Diosi:89}. Therefore, even though the K-model and D-model we consider here do not describe any real collapse of the wave function, they can still be used to get information about some relevant quantity, like the typical length and time scale over which coherence effects are suppressed, which would be the same also for the corresponding gravity induced collapse equations.

While both the models study gravity induced decoherence, and have some common features, they arrive at results which are quantitatively different at times. The purpose of the present paper is to compare and contrast the K-model and the D-model, and to understand why their results differ quantitatively. 

The K-model begins by asking to what precision a length $s=cT$ in flat spacetime can be measured using a quantum probe which obeys the uncertainty principle. Karolyhazy shows that there will be a minimum uncertainty $\Delta s$ in the inferred length which is given by the relation
\begin{equation}
{\Delta s}^3 \sim l_p^2 \; s,
\label{kuncertain}
\end{equation}
where $l_p =\sqrt{\hbar G/c^3}$ is the Planck length. This uncertainty is accounted for by hypothesizing that there coexist in spacetime a  family of curved metrics $(g_{\mu\nu})_{\beta}$, each of which yields a corresponding value $s_{\beta}$ for the measured length. The family of metrics is chosen in such a way that they average to $s=cT$ with a variance
\begin{equation}
\Delta s^2 = \langle(s-s_\beta)^2\rangle
\label{variance}
\end{equation}
(here $\langle...\rangle$ denotes the average over the metrics) which yields a $\Delta s$ which matches with the uncertainty given by (\ref{kuncertain}). As we will recall in the next section, this matching requires an appropriate choice for the family of metrics.

Given such a family of metrics, one studies the propagation of an initial wavefunction $\Psi_0$ for an object of mass $m$ and size $R$ (assuming a spherical shape), in different metrics $(g_{\mu\nu})_{\beta}$. Inevitably, the wave function $\Psi_\beta$ at a later time will belong to a set
$\{\Psi_\beta\}$ whose elements will differ from each other in their spatial dependence. In particular, the phase separation between two spatial points acquires a variance when averaged over the family, and a decoherence length scale $a_c$ (to which the wave function gets localized) is defined as the spatial separation $a_c$ over which the phase uncertainty becomes of the order of $\pi$. The decoherence time is given by $\tau_c \approx ma_c^2/\hbar$. 

Of great interest is the calculated dependence of the coherence length on the mass $m$ and the size $R$ of the object. The results of the K-model are as follows \cite{Frenkel:2002}. For an extended object of size $R$ the localization length is calculated from a generic form of the phase variance for a system of particles. There are two interesting cases: one for $R \gg a_c$ and another for $R \ll a_c$. For a micro-object of linear size $R\ll a_c$, the expression for the coherence length virtually reduces to that of an elementary particle of mass $m$ and the coherence length $a_c$ over which decoherence effects become relevant is given by
\begin{equation}
a_c \approx \frac{\hbar^2}{G}\; \frac{1}{m^3}  = \left(\frac{L}{l_p}\right)^{2} L; \qquad L= \frac{\hbar}{mc}.
\label{micro}
\end{equation} 
While for $R \gg a_c$, the critical length can be expressed as,
\begin{equation}
a_c \approx \left(\frac{\hbar^2}{G}\right)^{1/3}\; \frac{R^{2/3}}{m} = 
\left(\frac{R}{l_p}\right)^{2/3} L.
\label{macro}
\end{equation}
[Putting $a_c=R$ in Eqn. (\ref{macro}), the two expressions coincide. Thus $a_c=R$ denotes the transition from macro-regime to micro-regime.] Summarizing this, the following important inferences can be drawn:
\begin{eqnarray}  
a_c\gg R &\implies& \frac{\hbar^2}{G} \gg m^3 R \qquad: micro-region,\\
\nonumber\\
a_c \approx R &\implies& \frac{\hbar^2}{G} \approx m^3 R \qquad: transition-region,\\
\nonumber\\
a_c\ll R &\implies& \frac{\hbar^2}{G} \ll m^3 R \qquad: macro-region.
\end{eqnarray}

Some estimates are of interest.  From Eqn. (\ref{micro}) it can be estimated that for a proton
\begin{equation}
a_c \approx 10^{25} \ {\rm cm}, \qquad \tau_c \approx 10^{53} \ {\rm sec}.
\label{proton}
\end{equation}
What this means is that while according to quantum theory, an initial wave function for the proton would continue to spread indefinitely and forever, gravity induced decoherence causes its loss of coherence, after an enormous time of $10^{53}$ sec, to a very large length scale $10^{25}$ cm. Given these length and time scales, we are of course completely justified in thinking of the proton as a quantum mechanical object. Contrast this with a macroscopic ball of radius $R$ = 1 cm, having density 1 gm/cm$^3$, for which we get
\begin{equation}
a_c \approx 10^{-16} \ {\rm cm}, \qquad \tau_c \approx 10^{-4} \ {\rm sec}.
\label{ball}
\end{equation}
Furthermore, the transition from the micro to the macro domain occurs for $a_c = R$, which for a density of 1 gm/cm$^3$, works out from (\ref{macro}) to be
  \begin{equation}
a_{tr} \approx 10^{-5} \ {\rm cm}, \qquad \tau_{tr} \approx 10^{3} \ {\rm sec}, \qquad m_{tr} \approx 10^{-14} \ {\rm gm}.
\label{ball2}
\end{equation}
Notice that the coherence length $a_{tr}$ for transition matches with the favoured value for $r_c$ in the collapse models. The transition mass corresponds to about $10^{10}$ amu, which is still about five orders of magnitude higher than the largest masses (about $10^{5}$ amu) for which quantum position superposition has been observed through interferometry. 

The similarity between CSL and gravity induced localization has been discussed for instance in \cite{RMP:2012}. It is important though to emphasize that while CSL explains localization, as well as  realization of a specific random outcome upon measurement in accordance with the Born rule,  gravity models only explain localization [without selection of one outcome]. In this sense, these models should perhaps better be called models of gravity induced decoherence, instead of gravity induced collapse.

The next model of gravity induced collapse was developed by Di\'osi, and in a spirit somewhat similar to the K-model, the work begins by asking to what accuracy a Newtonian gravitational field ${\bf g}$ can be
measured by a quantum probe obeying the uncertainty principle. It is shown that the uncertainty 
$\delta {\bf g}$ in the measured field, averaged over a spacetime volume $VT$, is bounded by
\begin{equation}
(\delta\tilde{{\bf g}})^2 \geq G \hbar/VT\; .
\label{dbound}
\end{equation}
This result is in spirit similar to the Karolyhazy bound (\ref{kuncertain}) mentioned above, and later in this paper we will discuss the relation between these two bounds. Di\'osi models this uncertainty by introducing a classical stochastic potential, whose two  point correlation reflects this bound. The stochastic potential is then introduced as a source potential in the Schr\"{o}dinger equation, making the evolution of the wave function stochastic. A deterministic Markovian master equation can be deduced for the density matrix,  and the stochastic potential is responsible for decoherence of the density matrix. The decoherence time $\tau_c$ and the localization length $a_c$ (related as before by $\tau_c = ma_c^2/\hbar$) can be calculated. For a spherical object of mass $m$ and size $R$ the localization length is given in two limiting cases by
\begin{eqnarray}
a_{c}&\sim& (\hbar^2/Gm^3)^{1/4}R^{3/4}, \quad {\rm if} \quad Rm^3 \gg \hbar^2/G,\nonumber\\
&\sim& (\hbar^2/Gm^3)^{1/2}R^{1/2}, \quad {\rm if} \quad Rm^3 \ll \hbar^2/G\; .
\label{modelcoh}
\end{eqnarray}
The result in the first line, which is for the macro limit, should be compared with the corresponding K-model result, given by Eqn. (\ref{macro}). The two results are different, and one would like to understand the reasons for the difference. In the subsequent sections we compare the two models and show that while the two spacetime bounds (\ref{kuncertain}) and (\ref{dbound}) are equivalent, they do not imply a unique two-point noise correlation for the assumed stochastic potential. While the D-model assumes white noise, the noise correlation in the K-model is not white, and the corresponding master equation is non-Markovian. This confirms the earlier finding of \cite{DL1989} about the different noise in the two models; however unlike \cite{DL1989} we suggest that the minimal spacetime bounds in the two models are equivalent, and that the bound does not determine the noise correlation.

Quantitatively too, the two models show a major difference, even though in both models the micro to macro transition takes place at the same value of the coherence length: $a_{tr} = R = \hbar^2/Gm^3$. In the D-model, for a proton, assumed to have a classical radius $R\approx 10^{-13}$ cm, the localization length and the decoherence time are found to be $10^{6}$ cm and $10^{15}$ sec respectively, much smaller than the corresponding numbers ($10^{25}$ cm and $10^{53}$ sec) for the K-model. In the macro limit, the D-model gives a localization length $10^{-12}$ cm for $R=1$ cm and a density of 1 gm/cm$^3$, which is larger than the K-model value by four orders of magnitude. Strangely enough, this corresponds to a decoherence time of about $10^3$ sec, which is unreasonably high, in contrast to the more plausible value $10^{-4}$ sec yielded by the K-model. Considering that these numerical values are now of interest to experimentalists in the field, it is highly desirable to try and make unique model predictions which do not differ by many orders.

It should be emphasised that both in the K-model and in the D-model, the classical stochastic potential is {\it postulated} by way of an assumption, to represent the quantum space-time uncertainty. Consequently, strictly speaking, the two-point correlation functions associated with the corresponding stochastic fields are also  postulates, albeit ones that are motivated by certain physical and mathematical choices made in the respective models.

 The plan of this paper is as follows. In Section II we briefly review the K-model, and argue that one can think of the family of metrics as a stochastic potential. We then reconfirm the result of \cite{DL1989} on the two point noise correlation for this potential, and show that it corresponds to non-white noise. We also write down the non-Markovian master equation for this model.  In Section III we recall the D-model, and show how can one think of the uncertainty bounds of the two models, Eqns. (\ref{kuncertain}) and (\ref{dbound}) as being equivalent to each other. We also argue that the methods used by the two models for calculating the decoherence time are equivalent. In Section IV we highlight that the spacetime uncertainty bound does not uniquely determine the noise correlation, and some additional criterion will have to be sought to obtain a unique prediction for the localization length.  

\section{A brief review of the K-model, and some new results}

In the first part of this section we report some of the most important properties of the K-model, following the derivation given in \cite{Karolyhazi:86,Karolyhazi:66}.

As noted above, the bound (\ref{kuncertain}) is modeled by introducing a family of metrics $(g_{\mu\nu})_{\beta}$ which are very close to the Minkowski metric. $\beta=0$ labels the Minkowski metric. The proper length $s=cT$
 between two world-points ${\bf x_1}$ and ${\bf x_2}$ is defined as the mean of the lengths 
$s_\beta$ as measured in different members of the family
\begin{equation}
s=cT=\langle {s}_\beta \rangle
\label{avlength}
\end{equation}
and the uncertainty $\Delta s$ in the length of the line segment is given by Eqn. (\ref{variance}). Assuming the particle motion to be nonrelativistic, only the departure of the $g_{00}$ metric component from its Minkowski value is of interest, and one introduces the perturbation
\begin{equation}
(g_{00})_{\beta}({\bf x},t) = 1 + \gamma_{\beta}({\bf x},t)\; .
\label{defbeta}
\end{equation}
Now, the idea is to select the set $\gamma_\beta$ in such a way that the length of the world line
\begin{equation}
 s_{\beta}=\int dt\left[g_{\mu\nu}^{\beta}\frac{dx^{\mu}}{dt}\frac{dx^{\nu}}{dt}\right]^{1/2}
\end{equation}
averages to (\ref{avlength}), and the uncertainty in length as defined by Eqn. (\ref{variance}) matches with the bound (\ref{kuncertain}). For this purpose, the K-model assumes (since spacetime is free apart from matter particles) that the $\gamma_\beta$ satisfy the wave equation
\begin{equation}
\Box \gamma_{\beta}({\bf x},t) = 0\; .
\label{box}
\end{equation}
In Section IV we will point out that this is not a unique choice for $\gamma_\beta$ and other choices can also yield (\ref{kuncertain}).

To proceed further, it is convenient to make a Fourier expansion of the $\gamma_{\beta}$ satisfying
(\ref{box}) with $\omega=|{\bf k}|c$
\begin{equation}
\gamma_{\beta}({\bf x},t)=\frac{1}{\sqrt{l^{3}}}\sum_{{\bf k}}\left[c_{\beta}({\bf k}) e^{i({\bf k}\cdot{\bf x}-\omega t)}+c.c\right]\; .
\label{fourier}
\end{equation}
Also, the K-model assumes $c_{\beta}({\bf k})=f(k) e^{i\alpha_{\beta}({\bf k})}$, where $\alpha$ is a random phase such that
\begin{equation}
 \langle c_{\beta}({\bf k})\rangle =0, \qquad \langle c_{\beta}^{2}({\bf k})\rangle =0, \qquad \langle c_{\beta}({\bf k})c^{*}_{\beta}({\bf k'})\rangle = \delta_{{\bf k},{\bf k'}}(f(k))^{2}.
 \label{cond}
 \end{equation}
 This is a simplifying assumption of the model; namely that the $c_{\beta}({\bf k})$ are independent stochastic variables  with zero mean, and a Gaussian probability distribution (see Eqn. 4 of \cite{Karolyhazy:90}).
 It is then shown that in order to recover (\ref{kuncertain}) the function $f(k)$ is given by
 \begin{equation}
 f(k)=l_p^{2/3}k^{-5/6}.
 \end{equation}
 
 Having determined the family $\gamma_\beta$ the next task is to determine the evolution of a given initial wave function $\psi_0$, and find out how the evolution depends on $\gamma_{\beta}$. Different metrics will result in different evolution, thus leading to a family of wave functions $\psi_\beta$, all of which in fact describe the same system. Decoherence results when there is a significant difference in the evolution as $\beta$ is varied: this is quantified as follows.
 
 We start from the Klein-Gordon equation
 \begin{equation}
 \frac{1}{\sqrt{-g_{\beta}}}\frac{\partial}{\partial x^{\mu}}(\sqrt{-g_{\beta}}g^{\mu\nu}_{\beta}\frac{\partial \phi}{\partial x^{\nu}})-
 \left(\frac{mc}{\hbar}\right)^2\phi=0,\nonumber
\end{equation}
 and takes its nonrelativistic limit to arrive at the Schr\"{o}dinger equation
    \begin{equation}
 i\hbar\frac{\partial}{\partial t}\psi_{\beta}=\left(-\frac{\hbar^2}{2m}\triangledown^2+V_{\beta}\right)\psi_{\beta}
 \label{nonrel}
\end{equation}
 where the perturbing potential $V_\beta$ is given by
 $$V_{\beta}({\bf x},t)=\frac{mc^2\gamma_{\beta}({\bf x},t)}{2}.$$ 
 The non-relativistic limit has been arrived at by first substituting the metric form\\
$diag(g_{00}, -1, -1, -1)$ in the Klein-Gordon equation and using the perturbative expansion  
(\ref{defbeta}) for $g_{00}$. Then, as is conventionally done, the state $\phi$ in the Klein-Gordon equation is written as $\phi\equiv e^{iS}$ and the function $S$ expanded as a power series in $c^2$: $S = c^2 S_0 + S_1 + c^{-2} S_2 + ..$. Substitution of this expansion in the Klein-Gordon equation, and comparison of terms at different orders in $c^2$ yields the non-relativistic Schr\"{o}dinger equation at order $c^0$ after the identification  $\psi_\beta\equiv e^{iS_1/\hbar}$. A more detailed discussion can be found for instance in 
 \cite{Kiefer1991}.
 
 An important remark is in order with regard to Eqn. (\ref{nonrel}). Since this equation is being treated as the non-relativistic limit  of a relativistic system, semiclassical Einstein equations imply that in principle one ought to consider a contribution to the potential from self-gravity, of the form $\nabla^2 V_{self} \propto |\psi|^2$. The latter self-interaction is precisely what is considered in the Schr\"{o}dinger-Newton [SN] equation; however the SN equation does not consider the effect of spacetime uncertainty that is being studied in the K-model / D-model. In a sense the SN equation is complimentary to the present study, although it has its own limitations \cite{Bahrami2014a}, and in particular does not incorporate quantum fluctuations of the mean self-gravity. In our view a compete treatment should simultaneously include both self-gravity and the effects of intrinsic spacetime uncertainty. To the best of our knowledge this has not been done, and we hope to investigate this in the future.

 Generalization of the above non-relativistic equation to the many-particle case is achieved by replacing the potential $V_\beta$ by
 \begin{eqnarray}
 U_{\beta}(\{{\bf X}\},t)=\sum_i\frac{m_ic^2\gamma_{\beta}({\bf x}_i,t)}{2},
\label{multi-potnal}
\end{eqnarray}
where $\{{\bf X}\}$ labels a point in configuration space: ${\{{\bf X}\}}=({\bf x_1}, {\bf x_2}, ....{\bf x_N})$.
Then the Schr\"{o}dinger equation becomes:
  \begin{equation}
 i\hbar\frac{\partial}{\partial t}\Psi_{\beta}(\{{\bf X}\},t)=\left(H+U_{\beta}(\{{\bf X}\},t)\right)\Psi_{\beta}.
 \label{multi}
\end{equation}

 To realize decoherence one starts with an initial wave function $\Psi_{0}(\{{\bf X}\},0)$, same for all the metrics
 $\{g^{\mu\nu}_{\beta}\}$. After evolution, different $\Psi_{\beta}(\{{\bf X}\},t)$ will become different. It can be shown that,  to a good approximation \cite{Karolyhazy:90}
\begin{equation}
 \Psi_{\beta}(\{{\bf X}\},t)\approx\Psi_{0}(\{{\bf X}\},t)e^{i\phi_{\beta}(\{{\bf X}\},t)},
\end{equation}
with
\begin{equation}
 \phi_{\beta}(\{{\bf X}\},t)=-\frac{1}{\hbar}\int_{0}^{t}dt' U_{\beta}(\{{\bf X}\},t).
\end{equation}
We fix an $\{{\bf X_1}\}$ and an $\{{\bf X_2}\}$, and calculate the difference in phase between these two points in configuration space 
for different $\beta$. The answer will depend on $\beta$ and on time. The root mean square spread in the phase (average is over $\beta$)
$$\langle[\phi_{\beta}(\{{\bf X_1}\},t)-\phi_{\beta}(\{{\bf X_2}\},t)]^{2}\rangle^{1/2}$$
can be estimated as a function of $\{{\bf X_1}\}$, $\{{\bf X_2}\}$ and time $t$. The uncertainty in the relative phase depends only on the separation 
between the two points in configuration space, and  for a sufficiently large separation $a_c$ can reach the value $\pi$ after some time. When that happens, decoherence and localization is said to occur, and the aforementioned results (\ref{micro}) and (\ref{macro}) are shown to hold. [The phase correlations are assumed to be Gaussian, so that the two-point function carries the entire information about the correlations.] 

In our paper, we will attempt to recast the analysis of the K-model in a manner which might be regarded as more conventional, and which facilitates comparison with the D-model. Thus, there is nothing which really prevents us from thinking of the family $\gamma_{\beta}$ as a stochastic potential with zero mean, and whose two point correlation is such that when a length $s=cT$ is measured in the presence of such a potential, it exhibits an uncertainty given by (\ref{kuncertain}). In order to avoid possible divergences due to taking a discrete  distribution of point-like particles, we rewrite Eqn. (\ref{multi}) considering a system with mass density given by a smooth function $f({\bf x})$,
 \begin{equation}
 i\hbar\frac{\partial}{\partial t}\psi_{\beta}({\bf x}, t)=\left(H+\frac{c^2}{2}\int d^3 x^\prime \; 
 f({\bf x}^\prime-{\bf x}) \; \gamma_{\beta} ({\bf x}^\prime,t)\right)\psi_{\beta}({\bf x}, t),
 \label{multi2}
 \end{equation}
where $\gamma_{\beta}$ is now a stochastic potential, and the wave function is also a stochastic quantity, which represents the family $\psi_{\beta}$.

We now compute the two point correlation (assuming a Gaussian probability distribution) for the stochastic potential $\gamma_{\beta}$ in the K-model. By means of the Fourier expansion (\ref{fourier}) we can write,
\begin{equation}
\corr{\gamma_{\beta}({\bf x},t)}{\gamma_{\beta}({\bf x'},t')}=\frac{1}{l^{3}}\corr{\sum_{{\bf k}}\sum_{{\bf k'}}\left[c_{\beta}({\bf k}) e^{i({\bf k}\cdot{\bf x}-\omega t)}+c.c\right]}{\left[c_{\beta}({\bf k'}) e^{i({\bf k'}\cdot{\bf x'}-\omega' t')}+c.c\right]}
\end{equation}
and using relations (\ref{cond}) we obtain
\begin{equation}
\corr{\gamma_{\beta}({\bf x},t)}{\gamma_{\beta}({\bf x'},t')}=\frac{1}{l^3}\sum_{{\bf k}}\left[f^2(k) e^{i{\bf k}\cdot({\bf x}-{\bf x'})}e^{-i\omega(t-t')}+f^2(k) e^{-i{\bf k}\cdot({\bf x}-{\bf x'})}e^{i\omega(t-t')}\right].
\end{equation}
In the limit $l \to \infty$ we can write
\begin{equation}
\corr{\gamma_{\beta}({\bf x},t)}{\gamma_{\beta}({\bf x'},t')}=\frac{1}{(2\pi)^3}\int d{\bf k}\left[f^2(k) e^{i{\bf k}\cdot({\bf x}-{\bf x'})}e^{-i\omega(t-t')}+f^2(k) e^{-i{\bf k}\cdot({\bf x}-{\bf x'})}e^{i\omega(t-t')}\right]
\end{equation}
which, introducing $r=\vert{\bf x}-{\bf x'}\vert$ and $\tau=t-t'$ becomes
\begin{equation}
\corr{\gamma_{\beta}({\bf x},t)}{\gamma_{\beta}({\bf x'},t')}=\frac{1}{(2\pi)^3}\int d{\bf k}\left[f^2(k) e^{ikr\cos\theta}e^{-i\omega\tau}+f^2(k) e^{-ikr\cos\theta}e^{i\omega\tau}\right].
\end{equation}
Let us first calculate the first term:
\begin{eqnarray}
I_1&=&\frac{1}{(2\pi)^2}\int^{\infty}_{0}\int^{\pi}_{0}k^2 dk f^{2}(k)\sin\theta d\theta e^{ikr\cos\theta}e^{-i\omega\tau}\\
\nonumber\\
&=&\frac{l_p^{4/3}}{(2\pi)^2}\frac{2}{r}\int^{\infty}_{0}k^{-2/3} \sin(kr)e^{-ikc\tau} dk. \nonumber
\end{eqnarray}
Similarly, the second term gives,
\begin{equation}
I_2=\frac{l_p^{4/3}}{(2\pi)^2}\frac{2}{r}\int^{\infty}_{0}k^{-2/3} \sin(kr)e^{ikc\tau} dk 
\end{equation}
and, adding these two terms, we finally get:
\begin{equation}
\corr{\gamma_{\beta}({\bf x},t)}{\gamma_{\beta}({\bf x'},t')}=\frac{l_p^{4/3}}{(2\pi)^2}\frac{4}{r}\int^{\infty}_{0}k^{-2/3} \sin(kr)\cos(kc\tau) dk.
\end{equation}
Upon integration, the final form of two point correlation is,
\begin{equation}
\corr{\gamma_{\beta}({\bf x},t)}{\gamma_{\beta}({\bf x'},t')}= \frac{l_{p}^{4/3}}{4\pi^{2}r}\Gamma\left(1/3\right)\left[\frac{1}{(r+c|\tau|)^{\frac{1}{3}}}+\frac{\textrm{sign}(r-c|\tau|)}{|r-c|\tau||^{\frac{1}{3}}}\right].
\label{kcorr}
\end{equation}
This result was first reported in \cite{DL1989}, and 
this is evidently not white noise, and is the feature responsible for the difference in the results obtained for localization length in the K-model and the D-model. However, unlike what \cite{DL1989} seems to suggest, we will demonstrate in the next section  that the uncertainty bounds (\ref{kuncertain}) and (\ref{dbound}) are equivalent. Furthermore, the phase variance method used in the K-model to determine the decoherence time will be shown to be equivalent to the more conventional method used in the D-model (i.e. studying the master equation for the density matrix). Thus these are not the reasons why the two models arrive at different results.

We can also write down the non-Markovian master equation corresponding to this non-white noise. Rewriting (\ref{multi2}) without projecting to the position basis we have
\begin{equation}
i\hbar\frac{\partial}{\partial t}\left|\psi\left(t\right)\right\rangle =\left[H+\frac{c^{2}}{2}\int d^3 x' f\left({\bf x'}- \hat{{\bf q}}\right)\gamma_{\beta}\left({\bf x'},t\right)\right]\left|\psi\left(t\right)\right\rangle. \label{eq:single gen-2}
\end{equation}
It has been shown in \cite{Adler:2007}, that when the noise can be treated as a perturbation, the corresponding master equation to the lowest perturbative order in the noise is 
\begin{equation}
\frac{d\rho\left(t\right)}{dt}=-\frac{i}{\hbar}\left[H,\rho\left(t\right)\right]-\left(\frac{c^{2}}{2\hbar}\right)^{2}\int d^3 x d^3 x'\int_{0}^{t}dt' \corr{\gamma_{\beta}({\bf x},t)}{\gamma_{\beta}({\bf x'},t')} \left[f\left({\bf x}\right),\left[f\left({\bf x}'\left(t'-t\right)\right),\rho\left(t\right)\right]\right],
\end{equation}
with $f\left({\bf x}\right)=f\left({\bf x'}-\hat{{\bf q}}\right)$, $\corr{\gamma_{\beta}({\bf x},t)}{\gamma_{\beta}({\bf x'},t')}$ given in Eq. (\ref{kcorr}) and ${\bf x'}\left(t'-t\right)$ the position operator in interaction picture evolved up to the time $t'-t$. The calculation of decoherence time from this master equation is by no means straightforward nor obvious, and the phase variance method used in the K-model is decidedly far simpler. As we show in the next section, for the D-model, the phase variance method is equivalent to the Markovian master equation, in so far as the calculation of the decoherence time is concerned. We conjecture that the same equivalence is true, if that master equation is replaced by the above non-Markovian equation of the K-model, and that this latter equation also yields decoherence in position. However we do not have a proof for this, or a derivation of the phase variance method from this master equation, and we hope to address these questions in a future study.

\section{A brief review of the D-model, and comparison with the K-model}

Di\'osi and Luk\'acs \cite{Diosi:87a} work in the framework of Newtonian Quantum Gravity [NQG], where $G$ and $\hbar$ appear in the analysis, but $c$ does not. They ask: if a quantum probe is used to measure a classical gravitational field $\bf g$, what is the maximum accuracy with which ${\bf g}$ can be measured? They assume that in a realistic measurement only an average $\bf \tilde{g}$ of ${\bf g}$ 
\begin{equation}
{\bf \tilde{g}}({\bf x}, t) = \frac{1}{VT} \int {\bf g}({\bf x^\prime}, t) d^{3}x^\prime dt, 
\qquad |{\bf x}-{\bf x^\prime}| < R, \qquad |t^\prime -t| < \frac{T}{2}
\end{equation}
over a space and time interval can be measured. The volume $V=4\pi R^3/3$ and the time-interval $T$ are properties of the probe, assumed to be a spherical object with linear extent $R$.

The quantum probe, assumed to be a mass $M$ with  wave-packet of initial extent $R$ picks up a momentum 
${\bf P}=M{\bf \tilde{g}}T$ during the measurement time $T$. However there is a quantum uncertainty 
$\delta P \sim \hbar / R$ in the classical value of $P$, showing that there is an inaccuracy in the measurement of ${\bf \tilde{g}}$ given by $\delta ({\bf \tilde{g}}) \sim \hbar / MRT$. This inaccuracy can be decreased by increasing $M$, but the mass $M$ produces its own gravitational field, which disturbs the field being measured, and has an intrinsic uncertainty $\delta {\bf g}\sim GM/R^2$ because of the spread of the wave-packet. Consequently the optimal choice for $M$ is $M\sim \sqrt{\hbar R/GT}$ and the final minimal uncertainty in the measurement of the gravitational field is 
\begin{equation}
\delta(\tilde{\bf g}) \sim \sqrt{G\hbar/VT}. 
\label{minimal}
\end{equation}
This minimal uncertainty appears to have a universal character, and could be mathematically modeled, for the sake of further application, as a stochastic contribution ${\bf g}_{st}({\bf x}, t)$ (having zero mean) to the classical field ${\bf g}_{cl}({\bf x}, t)$
\begin{equation}
{\bf g}({\bf x}, t) = {\bf g}_{cl}({\bf x}, t) + {\bf g}_{st}({\bf x}, t)\; .
\end{equation}
The minimal bound (\ref{minimal}) can be recovered using this stochastic field, provided that, after spacetime averaging, it satisfies
\begin{equation}
\langle{(\tilde{{\bf g}}}_{st})^2\rangle \sim \frac{G\hbar}{VT}\; .
\label{mean}
\end{equation}
The spacetime average that appears on the left hand side can be written explicitly, so that
\begin{equation}
\langle{\bf \tilde{g}}_{st}^2\rangle = \frac{1}{V^2 T^2}\int \int d^3 x \; d^3 x^\prime \; dt \; dt^\prime \langle{\bf g}_{st}({\bf x}, t)
{\bf g}_{st} ({\bf x^\prime}, t^\prime)\rangle\; .
\label{avg}
\end{equation}
Now, as is easily verified, a possible two-point correlation on the right hand side which will satisfy this relation is white-noise:
\begin{equation}
\langle{\bf {g}_{st}({\bf x}, t})
{\bf g}_{st} ({\bf x^\prime}, t^\prime)\rangle = {\bf I} G\hbar \delta({\bf x} - {\bf x^\prime}) \delta (t - t^\prime)
\end{equation}
The form of the correlation leads to the cancellation of a factor $VT$ in (\ref{avg}), which is what is desired. From here it can be shown that the two point correlation for the gravitational potential $\phi$ defined by ${\bf g}=-\nabla\phi$ is given by
\begin{equation}
\langle\phi({\bf x}, t)\phi({\bf x}^\prime,t^\prime)\rangle - \langle\phi({\bf x}, t)\rangle\langle\phi({\bf x}^\prime,t^\prime)\rangle \sim \frac{G\hbar}{|{\bf x}-{\bf x}^\prime|}\;\delta(t-t^{\prime}),
\label{phicorr}
\end{equation}
where $\langle\phi({\bf x},t)\rangle=\phi_{cl}({\bf x},t)$ and $\phi_{cl}({\bf x},t)$ satisfies the Poisson equation
\begin{equation}
\nabla^2\phi({\bf x},t) = 4\pi G\rho({\bf x},t).
\end{equation}
As before, the probability distribution  for the stochastic potential is assumed to be Gaussian. 
From here on, the analysis proceeds in a straightforward manner: one uses the correlation function for the stochastic potential, and then constructs a Schr\"{o}dinger equation for evolution of an object in this stochastic potential. From there one can construct a Lindblad master equation, from which the decoherence effects of the fluctuating  gravitational potential can be deduced. This leads to Diosi's results mentioned above in Eqn. (\ref{modelcoh}).

Now, we notice that Eqns. (\ref{mean}) and (\ref{avg}) do not uniquely imply that the noise is white. For instance, suppose that in Eqn. (\ref{avg}) the correlation on the right hand side has a form
\begin{equation}
\langle{\bf g}_{st}({\bf x}, t)
{\bf g}_{st} ({\bf x^{\prime}}, t^{\prime})\rangle = {\bf I} G\hbar F({\bf x} - {\bf x^\prime})  G(t - t^{\prime})
\end{equation}
with $F$ and $G$ functions other than delta-functions. Now it is obvious, by defining new coordinates say, ${\bf p} = {\bf x} - {\bf x'}$, ${\bf q = {\bf x} + {\bf x'}}$, $r=t-t'$, $s=t+t'$, that the right hand side of Eqn. (\ref{avg}) can be written as
\begin{equation}
\frac{1}{VT} \int \; |det J|^{-1} \; d^3 p \; d r \; F({\bf p}) G({ r}).
\end{equation}
Here $J$ is the Jacobian of the transformation. This form can in principle yield (\ref{mean}) with a suitable choice of the functions $F$ and $G$ (other than delta functions), since a factor  $1/VT$ has again cancelled out. Thus it seems to us that the minimal bound (\ref{mean}) can also be achieved by noise which is not white, and this would make a difference in the final quantitative conclusions one draws about decoherence and how it depends on the mass and size of the decohering object. 

[For the sake of completeness we recall that the master equation for the D-model has an intrinsic divergence. To regularise this divergence a cut-off was first proposed at the nucleon scale \cite{Diosi:89}; unfortunately such a cut-off leads to excessive heating inconsistent with observations. It was then proposed to raise this cut-off to a much higher value $r_c$, the length scale of the CSL model \cite{Ghirardi3:90}. While this solves the heating problem, it is difficult to physically justify the inclusion of the scale $r_c$ in the D-model, one of whose motivations was to give a parameter free description of decoherence and collapse. In a recent paper \cite{Bassi2014}, the issue of this divergence has been discussed in detail, and the authors also consider the possibility of avoiding overheating by introducing dissipation in the dynamics, instead of introducing a high cut-off.]  

The second observation which we wish to make is the role played by spacetime averaging. In principle, it is possible to consider an idealized quantum probe whose spread is small enough that the gravitational field maybe assumed uniform over the extent of the probe and over the duration of the measurement. In such a case, the analysis leading upto Eqn. (\ref{minimal}) can be again repeated, but now without referring to spacetime averaging, and the minimal uncertainty will be given by
\begin{equation}
\delta({\bf g}) \sim \sqrt{G\hbar/VT},
\label{minimal2}
\end{equation}
where no reference to averaging is made. It is possible to relate this bound to the spacetime bound in the K-model, as we will see in a moment. Spacetime averaging is of course essential in the D-model, for the purpose of deducing a white noise correlation, as is evident from Eqn. (\ref{avg}).

If spacetime averaging is not done, and we accept the bound (\ref{minimal2}), then this can be related to the minimal bound in the K-model, once we think of Newtonian gravity as an approximation to general relativity. The bound (\ref{minimal2}) is equivalent to an uncertainty in the measured gravitational potential, given by
\begin{equation}
\delta(\phi)  \sim \sqrt{G\hbar/RT}\; .
\label{Dvariance}
\end{equation}
If the mean value of $\phi$ is zero, then $\delta (\phi)$ is of the order of the perturbing potential which distorts a Minkowski spacetime background. Now, if one attempts to measure a length $s=cT$, an uncertainty in its measurement will be induced by $\delta (\phi)$, and given as follows:
\begin{equation}
s^\prime = \sqrt{g_{00}} cT = \sqrt{1+ \frac{2\phi}{c^2}} cT \sim cT + \sqrt{\frac{G\hbar}{c^4 RT}}cT
\end{equation}
giving that
\begin{equation}
(\Delta s)^2 \equiv (s'- cT)^2 \sim l_p^2 \frac{s}{R},
\label{s1eqn}
\end{equation}
which with the assumption $R=\Delta s$ becomes the Karolyhazy relation
${\Delta s}^3 = l_p^2 \; s$. [Equating $R$ and $\Delta s$ is justified because we are looking for the minimum length uncertainty; a smaller value of $R$ would increase $\Delta s$ and a larger value of $R$ would imply that the actual uncertainty is $R$ and not $\Delta s$.]

Furthermore, it can be shown that the averaged potential in the D-model also implies the K-model spacetime bound, provided the white noise correlation is assumed.
We calculate the stochastic world line length similar to $s_{\beta}$ in K-model but now with a spacetime averaged potential $\tilde{\phi}=\frac{1}{VT}\int\int \phi({\bf x},t) \,d^{3}x \,dt$ :
\begin{equation}
s = \int_{0}^{T} \sqrt{1+\frac{2\tilde{\phi}}{c^{2}}} c \,dt.
\end{equation}
Note that here $\tilde{\phi}$ is a stochastic variable and the line element $s$ thus obtained is also stochastic.\\
Hence,
\begin{equation}
s-s_{avrg} \simeq c\int_{0}^{T}\frac{\tilde{\phi}}{c^{2}} \,dt
\end{equation}
\\where $s_{avrg}=cT$.\\
Next we calculate $\Delta s^{2} =\langle (s-s_{avrg})^{2}\rangle$:
$$\Delta s^{2}=\frac{1}{c^{2}}\corr{\int_{0}^{T}\tilde{\phi} \,dt'}{\int_{0}^{T}\tilde{\phi} \,dt''}.$$
As $\tilde{\phi}$ has already been averaged over the measuring time $T$, it depends only weakly on $t$ and can be thought to remain almost constant within time $T$. Using this, we can directly integrate over time to get
$$\Delta s^{2}= \frac{T^{2}}{c^{2}}\langle\tilde{\phi}^{2}\rangle \; .$$
Now, we explicitly calculate $\langle\tilde{\phi}^{2}\rangle $,
$$ \langle\tilde{\phi}^{2}\rangle = \frac{1}{V^{2}T^{2}}\corr{\int\int \phi({\bf x},t) \,d^{3}x \,dt}{\int\int \phi({\bf x'},t') \,d^{3}x' \,dt'}\; . $$
Using the two point correlation function (\ref{phicorr}) we get,
$$ \langle\tilde{\phi}^{2}\rangle = \frac{\hbar GT}{V^{2}T^{2}}\int\int \frac{1}{\vert\vec{x}-\vec{x'}\vert} \,d^{3}x \,d^{3}x'\,. $$
We now calculate the double integral by first denoting
$\vec{x}\equiv (x_{1}, x_{2}, x_{3}) $ and $ \vec{x'}\equiv (x'_{1}, x'_{2}, x'_{3}) $ and make a coordinate transformation  as follows:
$$ s= x_{1}-x'_{1}, \quad 
 p=x_{2} -x'_{2}, \quad
 q= x_{3}-x'_{3}, \quad
l = x_{1} + x'_{1}, \quad
m = x_{2} + x'_{2}, \quad
 n = x_{3} + x'_{3}. $$
This allows us to write the integral as:
$$\frac{1}{\vert\det J \vert}\int\int\frac{1}{\sqrt{s^{2}+p^{2}+q^{2}}} \,ds \,dp \,dq \,dl \,dm \,dn
= \frac{1}{\vert\det J \vert}\int\int\frac{1}{r} \,d^{3}r \,dl \,dm \,dn.$$
To find the limits of integration, we note that
since the dimension of the probe is $R$, we can say,
$$ x_{1},x_{2}, x_{3}, x'_{1}, x'_{2}, x'_{3} : -R/2 \to R/2, \qquad
 r : 0 \to R $$

and, evaluating the integral, we get
\begin{equation}
\int\int \frac{1}{\vert\vec{x}-\vec{x'}\vert} \,d^{3}x \,d^{3}x' \simeq  R^{5}. 
\end{equation}

This is of the order  $V^{2}/R $, and substituting
 this result back in the original equation, we get
\begin{equation}
\Delta s^{2}=\frac{ l_p^{2}}{R} s,
\label{seqn}
\end{equation}
where $s=cT$.
Now $R \approx \Delta s $ reproduces the K-model bound,
\begin{equation}
\Delta s^{3}= l_p^{2} s\; .
\end{equation}
We thus see that the minimal spacetime bounds in the D-model and in the K-model are essentially equivalent, and the difference in their final results is coming about because in one model the noise is white, and in the other it is not.

We have seen above that in the K-model the decoherence time is estimated by setting the phase variance to be of the order $\pi^2$. On the other hand, in the D-model, decoherence time is estimated from the master equation. We show below that in the D-model, the phase variance method gives the same result for decoherence time, as the master equation. This suggests that the phase variance method could well be sufficiently general, and equivalent to the master equation method, for a non-Markovian equation as well.

The phase of a wavefunction moving in a potential $U_{\beta}({\bf x},t)$ after time T is, $$\delta_{\beta}({\bf x},T)= -\frac{1}{\hbar}\int_{0}^{T}U_{\beta}({\bf x},t)\,dt\; .$$
In Di\'osi's model, the potential for a given configuration ${\bf X}$ is given as follows \cite{Diosi:87}:
\begin{equation}
U({\bf X},t)=\int_{vol} \phi({\bf x},t)f({\bf x}\vert {\bf X})\,d^{3}x
\end{equation}
where $f({\bf x}\vert {\bf X})$ denotes the mass density function at a point ${\bf x}$ for the configuration ${\bf X}$.\\

Thus  phases accumulated at time $t$ 
at configuration ${\bf X}$ is 
\begin{eqnarray}
\delta({\bf X},t)&=&-\frac{1}{\hbar}\int_{0}^{t}\int_{vol}\phi({\bf x'},t')f({\bf x'}\vert {\bf X})\, d^{3}x' \,dt'\,,
\end{eqnarray}
We now evaluate the variance $\langle[\delta({\bf X},t)-\delta({\bf X'},t)]^{2}\rangle$. We can write
$$ \langle[\delta({\bf X},t)-\delta({\bf X'},t)]^{2}\rangle = \frac{1}{\hbar^{2}}\left\langle\left[\int_{0}^{t}\int_{vol}\phi({\bf x'},t')f({\bf x'}\vert {\bf X})\, d^{3}x' \,dt' - \int_{0}^{t}\int_{vol}\phi({\bf x''},t'')f({\bf x''}\vert {\bf X'})\, d^{3}x'' \,dt''\right]^{2} \right\rangle\,,$$
calculate term by term and take the stochastic average using Eqn. (\ref{phicorr}).
The square modulus of the first term is
\begin{eqnarray}
\corr{\int_{0}^{T}\int_{vol}\phi({\bf x'},t')f({\bf x'}\vert {\bf X})}{\int_{0}^{T}\int_{vol}\phi({\bf x},t'')f({\bf x}\vert {\bf X})\, d^{3}x \,d^{3}x' \,dt' \,dt''} \nonumber\\
 = \frac{GT}{\hbar}\int_{vol}\int_{vol}f({\bf x'}\vert {\bf X})f({\bf x}\vert {\bf X})\frac{1}{\vert {\bf x}-{\bf x'}\vert}\,d^{3}x \,d^{3}x'. 
\end{eqnarray}
Similarly evaluating the square modulus of the second term 
and the cross term 
and summing those, we finally get,
\begin{equation}
\langle[\delta({\bf X},T)-\delta({\bf X'},T)]^{2}\rangle= \frac{GT}{\hbar}\int\int \,d^{3}x \,d^{3}x' \frac{[f({\bf x}\vert {\bf X})-f({\bf x}\vert {\bf X'})][f({\bf x'}\vert {\bf X})-f({\bf x'}\vert {\bf X'})]}{\vert {\bf x}-{\bf x'} \vert}\; .
\end{equation}
For decoherence, we need $\langle[\delta({\bf X},T)-\delta({\bf X'},T)]^{2}\rangle \approx \pi ^{2}$.
This gives us a decay time scale,
\begin{equation}
\tau_{d}^{-1} = \frac{G}{\pi^{2}\hbar}\int\int \,d^{3}x \,d^{3}x' \frac{[f({\bf x}\vert {\bf X})-f({\bf x}\vert {\bf X'})][f({\bf x'}\vert {\bf X})-f({\bf x'}\vert {\bf X'})]}{\vert {\bf x}-{\bf x'} \vert}.
\end{equation}
This is same as the decay time obtained by Di\'osi using the master equation apart from some constant factors. Thus use of  master equation or phase variance method gives similar result for the decay time. This suggests that the phase variance method is possibly a more general one which can give decay time for both white and non-white noise.

\section{Spacetime uncertainty bound and the noise correlation}

We have seen that in K-model and in D-model, the correlation functions of the potentials are completely different. 
Let us now try to find a general form of such potentials which would satisfy the following bound:
\begin{equation}
\Delta s^3 \sim l_p^2s.
\end{equation}  
The most general form of the potential is,
\begin{equation}
\phi({\bf x},t)=K F_{st}({\bf x},t),
\end{equation}
where $K$ is a constant with a suitable combination of $G$, $c$ and $\hbar$.
Now we will use this form to calculate the uncertainty in the length of a line element.
Then we have,
$$ s'=\int^T_0 \sqrt{g_{00}}cdt
= \int^T_0 \sqrt{1+K F_{st}({\bf x},t)}cdt
\approx c\int^T_0 \Big(1+\frac{1}{2}KF_{st}({\bf x},t)\Big)dt.$$
If we write $s=cT$, then
\begin{equation}\label{s'-s}
(s'-s)^2=\frac{K^2 c^2}{4}\Big(\int^T_0F_{st}({\bf x},t)dt\Big)^2\; .
\end{equation}
The uncertainty in the measurement of the length is obtained, as in the K-model, by averaging Eqn. (\ref{s'-s}): $\Delta s^2=\langle(s'-s)^2\rangle$.
We assume the correlation function of $F_{st}({\bf x},t)$ to be separable in space and time, i.e.
\begin{equation}
\langle F_{st}({\bf x},t)F_{st}({\bf x}',t') \rangle=P({\bf x},{\bf x}')g(t,t'),
\end{equation}
which leads to:
\begin{equation}
\Delta s^2=\frac{K^2 c^2}{4}P({\bf x},{\bf x})\int_0^T\int_0^T g(t,t') \,dt \,dt'\; .
\end{equation}
The linear size of the object under consideration is $R$.
At this point we impose the following three relations:
$$ \Delta s^3 \sim l_{p}^2 s \qquad
R \sim \Delta s \qquad
 s \sim cT .$$ 
For the above class of solutions, where we assumed that the noise correlation is separable in space and time coordinates, to obtain the Karolyhazy uncertainty relation, $P({\bf x},{\bf x})$ must be independent of ${\bf x}$ (e.g. $P({\bf x},{\bf x'})$ might depend on $|{\bf x}-{\bf x'}|) $, but it can be a function of $R$ and $g(t,t')$ can have a number of solutions such that the two point correlation function is given by:
\begin{equation}\label{g(t,t')}
g(t,t') =T^m t^{n_1} t^{'n_2}.
\end{equation}
We note that the form of the correlation function suggested above is neither the most general nor motivated from symmetries. However a large number of correlation functions can be constructed from the form given above and the linear combinations thereof. In general, the correlation function can take other forms also  (e.g. a function of $(t-t')$ as we have already shown in the previous section). 

We now show below how different solutions lead to the same K-bound.
Using the two point correlation in Eqn. (\ref{g(t,t')}), we find,
$$ \Delta s^2=\frac{K^2 c^2}{4}P({\bf x},{\bf x})\frac{T^{m+n_1+n_2+2}}{(n_1+1)(n_2+1)}.$$
The constants can be adjusted to reproduce the K-bound from the above equation for different choices of $\{m,n_1,n_2\}$. Below we illustrate some examples for a simplified version where $n_1=n_2=n$.

\begin{enumerate}
\item[1)] $P({\bf x},{\bf x})=\frac{1}{R}$.
In this case, we have,
$$\Delta s^2 \propto \frac{1}{R} T^{m+2n+2}.$$
Now, since $R \sim \Delta s $ and $ s \propto T $, we can write, from the above equation,
$$ \Delta s^3 \propto s^{m+2n+2}$$
and so, we must have, $m+2n+2=1$
which gives $m/2+n=-1/2 $.
For this, we find, 
\begin{equation}
\langle \phi^2({\bf x},t) \rangle =\frac{K^{2}}{R}T^m t^{2n}.
\end{equation}
 All such $\phi({\bf x},t)$, for which 
$\frac{m}{2}+n=-\frac{1}{2}$, are possible solutions.
Note that $m=-1 , n=0$ gives the form $\langle \phi^2 \rangle = \frac{G\hbar}{RT}$ which we had already predicted as a possible form of Di\'osi stochastic potential in Eqn. (\ref{Dvariance}).
\item[2)] $P({\bf x},{\bf x})=1$.
We get
$$ \Delta s^2 \propto s^{m+2n+2}.$$
Since, according to K-bound, it must be $$\Delta s^2 \propto s^{2/3},$$ we get the condition:
$$m+2n+2=\frac{2}{3}.$$

In general, we can say that if $P({\bf x},{\bf x})$ has a form $P({\bf x},{\bf x})=R^{2j}$ where $j$ is real, then the condition it must satisfy is
\begin{equation}
1-j=\frac{3}{2}(m+2n+2).
\end{equation}

\end{enumerate}
Thus, we see that for different choices of $j$, $m$ and $n$ we get different potentials all satisfying the Karolyhazy uncertainty relation. This has been shown for separable forms only. In general, the solution can be non-separable also, as in the K-model, where $\gamma({\bf x},t)$ cannot be separated in space and time coordinates. 

We conclude that, given the uncertainty in measurement or the space-time bound, the form of the potential cannot be uniquely determined. There is a whole class of solutions as discussed which lead to the same bound. It seems that the $\gamma$ in K-model and $\phi$ in Di\'osi's model are two special cases which simplify the mathematical treatment, but they are not unique choices. We are not suggesting that the above examples given by us are necessarily that of physically realistic noise, but rather that more work needs to be done to uniquely determine the gravitational noise correlation.

It is important to compare our analysis with that of Di\'osi and Luk\'acs [DL] \cite{DL1989} who in their Eqn. (8) propose that the fundamental geodesic uncertainty relation is
\begin{equation}
\Delta s^2 \approx l_p^2 \frac{s}{R},
\label{DL}
\end{equation}
which is the same as our Eqns. (\ref{s1eqn}) and (\ref{seqn}) but different from the Karolyhazy relation
(\ref{kuncertain}). To our understanding, DL suggest that (\ref{DL}), rather than (\ref{kuncertain}), is the fundamental relation. Through their analysis leading up to their Eqn. (11), which is
\begin{equation}
(\Phi)_{R,T}\approx\sqrt{\hbar G/RT},
\end{equation}
 they relate (\ref{DL}) to the metric uncertainty in their Eqn. (11), which is the same as our Eqn. (\ref{Dvariance}). On the other hand, as we have argued below Eqn. (\ref{s1eqn}), one should set $R\sim \Delta s$ for optimal minimal uncertainty, in which case Eqn. (\ref{DL}) above becomes the same as the Karolyhazy uncertainty bound (\ref{kuncertain}). In (\ref{DL}) above, it appears that increasing 
 $R$ decreases $\Delta s$, but if $R>\Delta s$ this does not seem physically reasonable, since the uncertainty would be bounded from below by the probe size $R$. Hence $R\sim \Delta s$ seems optimal.

Thus it seems to us that the main difference between the work of DL and our work is that whereas DL suggest the spacetime bounds in the two models are different, we have argued that the two bounds are actually equivalent to each other. Also, as we have attempted to demonstrate, the bound does not by itself favor white noise over colored noise, nor the other way around. It would appear that this issue is open for further examination.

\section{Summary and concluding remarks} 
The K-model proposes a minimal spacetime uncertainty bound, namely the accuracy with which a length interval can be measured by a quantum probe. This bound is realized via a hypothesized coexisting family of metrics. The propagation of the wave function in this kind of a hazy spacetime is shown to lead to loss of coherence, which becomes relevant for macroscopic objects. We showed that this family of metrics can equivalently be interpreted as a stochastic potential, whose two-point noise correlation can be worked out, and shown to be non-white noise.  The Schr\"{o}dinger evolution takes place in the presence of this stochastic potential, and the master equation for the density matrix is non-Markovian. 

The D-model also proposes an, apparently different, spacetime bound, i.e. the accuracy with which the gravitational field averaged over a spacetime region can be measured by a quantum probe. This uncertainty in the gravitational field can be modeled by a stochastic potential, which is assumed to have a white  noise correlation.  The Schr\"{o}dinger evolution of the wave function takes place in this stochastic potential, making the wave function stochastic.  A Markovian master equation for the density matrix is set up, and once again decoherence in position basis for large objects is demonstrated. The quantitative estimates for the decoherence time and localization length are however different from those of the K-model. 

We showed that the spacetime uncertainty bounds in the two models  are essentially equivalent to each other.  We then argued that the difference in the quantitative results of the two models is due to the assumed nature of the noise - white in one case, and coloured in the other case. We also argued that the spacetime bound does not uniquely predict the noise correlation, and many choices are possible, each of which is likely to give different results for the decoherence time scale. White noise may be the simplest choice, but there seems to be  no physical reason why gravitational effects must conform to white noise. Thus it would appear that additional criteria, apart from the minimal bound, are essential to precisely define a model of gravity induced decoherence.  Nonetheless, it can be said that the role of gravity in decoherence is fundamentally suggested, and further investigation of this problem is highly desirable. 

\bigskip

\noindent {\bf Acknowledgements:} The authors are grateful to Lajos Di\'osi for bringing Ref. \cite{DL1989} to their attention, and for helpful correspondence.The work of TPS is supported by Grant \# 39530  from the John Templeton Foundation. 
 SD acknowledges support from NANOQUESTFIT, the COST Action MP1006 and INFN, Italy. SD and KL acknowledge the hospitality of the Tata Institute of Fundamental Research (Mumbai) where part of this work has been done. TPS would like to thank Aniket Agrawal for collaboration during the early stages of this work.

\bigskip

\bigskip

\centerline{\bf REFERENCES}

\bibliography{biblioqmts3}

\end{document}